# What is User Experience Really: towards a UX Conceptual Framework

Stefan Hellweger and Xiaofeng Wang
Free University of Bolzano
Bolzano, Italy

*Abstract*—For more then a decade the term User Experience (UX) has been highly debated and defined in many ways. However, often UX remains as a vague concept and it may be hard to understand the very nature of it. In this paper we aimed at providing a better understanding of this concept. We explored the multi-faceted UX literature, reviewing the current state-of-the-art knowledge and emphasizing the multi-dimensional nature of the concept. Based on the literature review we built a conceptual framework of UX using the elements that are linked to it and reported in different studies. To show the potential use of the framework, we examined the UX delivered by different phone applications on different mobile devices using the elements in the framework. Several interesting insights have been obtained in terms of how the phone applications deliver different UX. Our study opens up a promising line of investigating the contemporary meaning of UX.

## I. INTRODUCTION

User experience (UX) has been a frequently discussed topic in software engineering literature in the past years. It is claimed that a paradigm shift from service to experience economy already happened in our time [1]. Like many other buzzwords, UX is defined differently and used to mean different things in different studies. No consensus has been reached on what UX means exactly. What Don Norman, the inventor of the term *user experience*, commented about 15 years ago is still valid today: "*I invented the term because I thought human interface and usability were too narrow. I wanted to cover all aspects of the person's experience with the system including industrial design, graphics, the interface, the physical interaction, and the manual. Since then the term has spread widely, so much so that it is starting to lose it's meaning... People use them often without having any idea why, what the word means, its origin, history, or what it's about.*"[1] This reflects the complex nature of UX.

The study presented in this paper is motivated by this observation. The research objective is to provide a better understanding of UX grounded in the existing literature. Rather than attempting to unify different definitions of UX artificially, we admit the multi-dimensional, multi-faceted nature of UX. Drawing upon the review of a set of studies that contain the definitions of UX, we propose a conceptual framework that may help improve the understanding of UX, both conceptually and practically.

The remaining of the paper is organized as below. Section II provides an overview of the understanding of UX in the literature. Section III presents the conceptual framework built upon the elements extracted from the reviewed literature. The framework is used in Section IV to analyze several phone applications in different mobile platforms from the UX perspective. The paper ends with a call for further research on conceptualizing UX.

## II. DEFINITIONS OF UX

Despite the fact that there is no consensus on the definition of UX in literature, there is a common understanding that it is a complex concept and should not be equaled to usability or user interface simply. Folstad and Rolfsen [2] contend that the literature on UX may be divided in three 'camps' in terms of the relation to usability: UX encompasses usability, UX complements usability, and UX is one of several components constituting usability. For example, Hassenzahl et al. [3] argue that, instead of merely making a software usable, an expanded perspective on usability would advance the designing of user experience. Being both usable and interesting, a software system might be regarded as appealing and as a consequence the user may enjoy using it. Stage [4] argues that the recent advent of systems are focusing more on amusement and entertainment and less on work in the traditional sense, which has led some to suggest a broader notion of usability with a significantly stronger focus on UX. Based on their previous work, Hassenzahl et al. [5] summarize important distinctions between the traditional view of usability and UX. They argue that UX takes a more holistic approach, aiming for a balance between pragmatic aspects and other non-task related aspects (hedonic) of product possession and use, such as beauty, challenge, stimulation, or self-expression. In addition, UX augments the "subjective." It is explicitly interested in the way people experience and judge products they use. What's more, UX is a more positive quality. Usability as a quality equals the removal of potential dissatisfaction. But even the best usability may never be able to "put a smile on users' faces." UX on the other hand addresses both, dissatisfiers and satisfiers, on an equal footing. The shift of emphasis from usability to experiential factors has forced researchers to consider what UX actually is and how to evaluate it [6].

Three dimensions of UX are most often suggested in the reviewed literature: user, product and interaction. As Forlizzi and Ford [7] suggest, a simple way to think about what influences experience is to think about the components of a user-product interaction, and what surrounds it. Arhippainen and Tähti [8] define UX as the experience that a person gets when he/she interacts with a product in particular conditions. The user and the product interact in the particular context of use that social and cultural factors are influencing. The user has the aspects including values, emotions, expectations and prior experience. The product has influential factors, for example, mobility and adaptivity. All these factors influence the experience that user-product interaction evokes. Similarly, Forlizzi and Battarbee [9] admit that understanding UX is complex. Designing the UX for interactive systems is even more complex, particularly when conducted by a team of multidisciplinary experts. They find that some approaches take the perspective of the user, others attempt to understand experience as it relates to the product, and a third group attempts to understand UX through the interaction between user and product. In one of the most cited UX papers, Hassenzahl and Tractinsky [10] emphasize again these three dimensions. They define UX as a consequence of a user's internal state (predispositions, expectations, needs, motivation, mood, etc.), the characteristics of the designed system (e.g. complexity, purpose, usability, functionality, etc.) and the context (or the environment) within which the interaction occurs (e.g. organizational/social setting, meaningfulness of the activity, voluntariness of use, etc.). Roto [11] takes the three components defined in [10] as a starting point and, with the knowledge on mobile browsing UX, identifies a set of attributes applicable for a wide range of UX cases.

---
[1] http://peterme.com/index112498.html

There are also other proposals in terms of the UX dimensions, even though much less dominant. For example, in [4] UX is redefined in terms of four factors where usability is one, and the others are: branding, functionality and content. It can be argued that this redefinition reflects a more product-focused approach to UX. Oygur and McCoy [12] suggest that UX is composed of tangible (e.g., physical needs, space requirements, ergonomic issues) and intangible (e.g., emotional needs, values) aspects.

UX can be approached in a more interdisciplinary manner [9]. There are quite diverse disciplines that enable different perspectives on UX. Broadly speaking the three main perspectives are IT, design and psychology. As observed by Vliet and Mulder [13], the discussion on human experience has a long (philosophical) tradition, further explored by psychologists, neurologist and others in the last centuries up until the current time. However this vast legacy of research on human experience has for a large part not found its way into current literature on Human-Computer Interaction, Interaction Design and Usability Engineering when addressing UX. Karapanos et al. [14] discuss two threads in the UX research. One has its roots in pragmatist philosophy and the other in social psychology. More and more studies emphasize on the non-instrumental aspect of UX and delve into understanding the physio, socio, psycho and ideo needs of human beings [15], [16].

III. CONSTRUCTION OF A UX CONCEPTUAL FRAMEWORK

Given the multi-disciplinary nature of UX, it is difficult to obtain all the relevant papers from different disciplines using automatic search engines. Therefore we used a more traditional, manual snowballing approach and gathered a collection of 21 papers (see the Reference list), which contain original definitions of UX. The majority of them come from Software Engineering related fields, some also from Design and Psychology. All the terms used to define UX in the papers were extracted and analyzed carefully. This resulted to 114 UX-related terms initially.

To group and present these items in a systematic manner was a challenge in this study. The three dimensions reported in Section II - User, Product and Interaction – turned out to be insufficient to cover the complexity and diversity shown by these terms. As a consequence we adopted a bottom-up approach to group the items. The emergent dimensions are Impacting Factors that affect UX, UX Characteristics and the Effects produced by UX. The resulting UX conceptual framework is presented in Fig.1.

IV. APPLICATION OF THE UX CONCEPTUAL FRAMEWORK

To illustrate how the conceptual framework can be helpful to understand UX practically, we chose a most basic feature of our modern smart phones to analyze: dialing a phone number to make a call. Even though almost a secondary feature nowadays, the dialing feature is still of vital importance for our mobile phones, at least the first time we call someone from our phone. Fig.2 shows the choice of mobile devices and the phone call feature built in them. These mobile phone brands were chosen based on their popularity and importance[2,3]. In addition to the built-in dialing features we also took Skype in the comparison, since it is one of the biggest voice and video communication providers with its users consuming daily 2 billion minutes in total[4].

Using the conceptual framework the first author, who is also a knowledgeable and experienced software developer, analyzed the user experience of these dialing features. The framework elements involved in the analysis are underlined and in *italic* in the following sub-sections. Note that the following examples serve the purpose to illustrate a possible application of the framework, and they are not intended to be exhaustive evaluation of the dialing features in these devices.

*A. The Dialing Pad*

Priori to iPhone nearly all mobile phones had a hardware keypad following the E.161 Standard or ISO 9995-8. The layout of the keypad was preserved in the various touch pads nowadays, as shown in Fig.2. The similar dialing pads across different phones reflected our *previous experience /memories* with mobile phones. The familiarity and resemblance to the legacy systems, phones in our case, ensure the *learnability* for everyone who previously used or saw a legacy system.

---

[2] http://www.statista.com/statistics/263401/global-apple-iphone-sales-since-3rd-quarter-2007/

[3] http://opensignal.com/reports/2014/android-fragmentation/)

[4] http://blogs.skype.com/2013/04/03/thanks-for-making-skype-a-part-of-your-daily-lives-2-billion-minutes-a-day/

*B. The "+" Button*

When we call abroad we need an international exit code for the actual call. These exit codes may be different depending on the country we are actually in. While in all the European Countries the code is 00, in the US the code is 011. In some countries the exit code even changes from operator to operator as for example in Colombia or Brazil. The "+" button was introduced as a placeholder suggesting it has to be replaced according to the correct country code. This feature increases the *effectiveness* and decreases the *complexity* of the dialing process.

The Skype dialing experience is consistent on Android and iOS devices, providing good *identification* and *esthetic* familiarity. The interesting difference from the other phone dial features is the implementation of the "+" feature, since it always displays the international prefix in the input field.

*C. The "Delete" Button*

From the introduction of iOS v1 in 2007 till v6 in 2012, the dial screen stayed the same. The look and feel as well as the screen size stayed unchanged. In the same time the design of the hardware (iPhone till iPhone5) was radically changed. The better use of the screen size in smartphones leads to a more pleasurable usage of the dial screen in smartphones. One subtle change is the "delete" button. In iOS before v5 (leftmost in Fig.2) it appears at the bottom right, close to the dial button (Galaxy S5 has a similar design). This design could cause a potential problem when we use the phone with one hand only (most of us are right handed), which is hitting the delete button accidentally while typing numbers.

In comparison, in the dialing pad in iOS since v5 (second left in Fig.2), the delete button is no more visible on the initial dialing pad. It only appears next to the input field when a number is digited, and therefore when the "delete" function is really needed (Nexus 5 has a similar design). This reflects the awareness of *task context*, and can be considered a more sensible design. In addition with a button less there is also a larger space for the dial keys, increasing the *usability* of the dialing feature.

*D. Design Aspects*

In iOS v7 the shape of the buttons was changed from a simple squared grid (leftmost in Fig.2) to round slightly space buttons (second left in Fig.2). This alignment with the hardware design improved the perception of the phone product as one. The hardware and the software are converging to provide a unified *esthetic experience*.

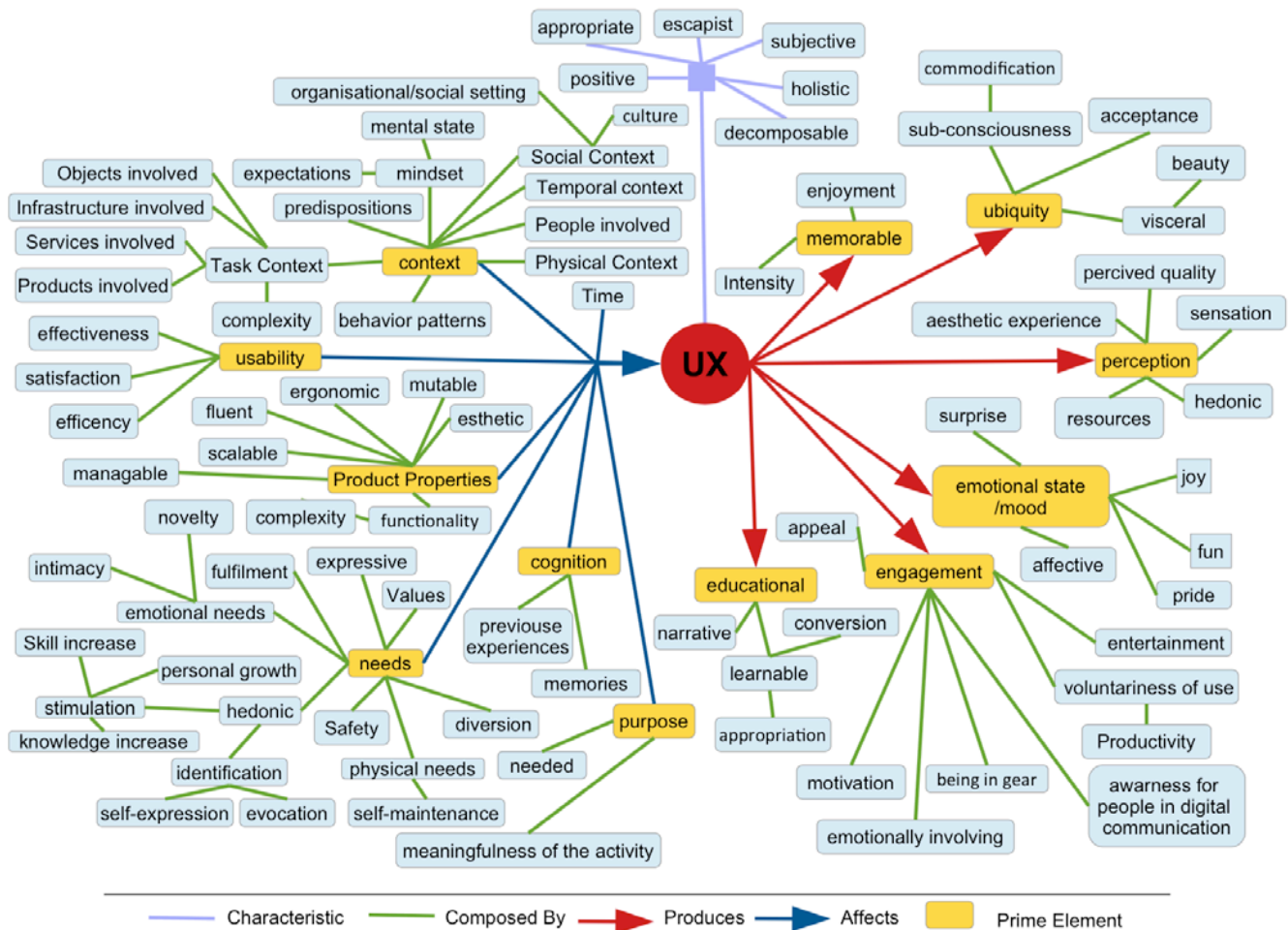

Fig.1 The Proposed UX Conceptual Framework

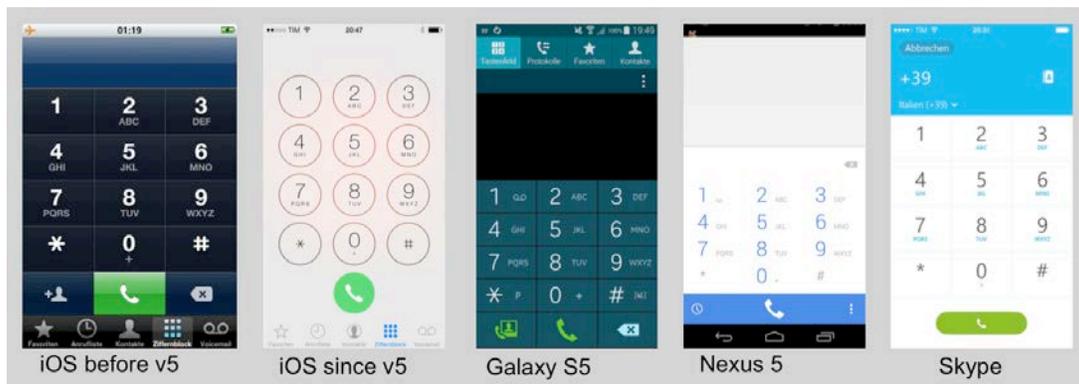

Fig.2 Different Phone Applications on Different Mobile Devices

## V. Discussion And Conclusion

The elements from the proposed UX conceptual framework helped to understand the evolution and design choices of the dialing features in different mobile devices. Some subtle changes, such as the example of the "delete" button, cannot be easily appreciated without the help of the UX elements. We contend that the proposed framework can help to increase the awareness of and sensibility to the UX provided by various software products and services.

It worth noting that the analysis presented in this paper is subjective and depends on the experience of the first author. Another limitation is that the analysis only applied the elements, not the relations among them, to make sense of the UX provided by these phone features.

Our study is still at the early stage. The framework needs to be refined. Future work also includes a systematic evaluation of its usefulness, e.g., different evaluators evaluate a given software product or service using the framework, and the results need to be compared systematically. In addition the application of the framework can be made more automatic and user friendly by building a UX evaluation tool on top of it.